**Rapid Stabilization of Droplets by Particles in Microfluidics: Role of Droplet Formation**

*Laura Andreina Chacon Orellana, and Jean-Christophe Baret\**

Dr. L. Chacon,
Univ. Bordeaux, CNRS, Centre de Recherche Paul Pascal, UMR5031, 33600 Pessac, France
Prof. J. -C. Baret
Univ. Bordeaux, CNRS, Centre de Recherche Paul Pascal, UMR5031, 33600 Pessac, France
and Institut Universitaire de France, France
E-mail: jean-christophe.baret@u-bordeaux.fr

Droplet-based microfluidics has emerged as a powerful technology for the miniaturization and automation of biochemical assays. The replacement of surfactants by nanoparticles as interfacial stabilizers has gained increasing interest. However, the stabilization mechanism of droplets by nanoparticles in microchannels is poorly understood, drastically hindering the development of practical applications. Current methods for droplet stabilization involve a trade-off between low droplet production throughput and waste of large number of nanoparticles. Here, we introduce a modification to the droplet production junction that reduces the droplet stabilization time by an order of magnitude, and at the same time significantly reduces the particle waste. Our results show that the limiting step in the kinetics of stabilization is the initial time where both phases come into contact and offer a guideline for the design of particle-stabilized droplet production devices.

## 1. Introduction

Droplet-based microfluidics is an efficient technology to generate, manipulate and analyze at ultra-high throughput droplets carried by an immiscible phase inside microchannels.[1] Droplets are used as miniaturized and isolated microreactors for controlled and high-fidelity (bio-)chemical reactions, with applications in single cell analysis and sequencing, cell selection and molecular diagnostics.[2-4]

Surfactants are commonly used for droplet stabilization in microfluidics, but particle-stabilized droplets –so-called Pickering emulsions– have been gaining increasing interest owing to their intrinsic interfacial properties: nanoparticles irreversibly adsorb to the droplet interface and provide a rigid interface preventing droplet break-up in narrow constrictions; they prevent cross-talk between droplets; they are chemically tunable allowing for the introduction of additional functionalities to the droplet interface opening potential applications with cell types which viability depends on their adhesion to a suitable substrate.[5-11]

However, Pickering emulsions introduce new challenges that need to be addressed to achieve a full optimization of these systems within droplet-based microfluidics technology. For instance, the surface chemistry of particles is critical as it controls several independent properties such as the emulsion stability, rheological properties and biocompatibility.[12-14]

From an engineering perspective, Pickering emulsions also have a limited throughput of production: the typical timescale required to stabilize them (~600 ms) is significantly higher than the needed for surfactant stabilized emulsions (~35 ms or less).[15-18]

The bottle-neck in the throughput lies in the stabilization time of the emulsion, which relates to the inability of the nanoparticles to quickly and efficiently reach the interface. A method typically used to reduce the droplet stabilization time scale consists of increasing the concentration of particles in the continuous phase to a large excess. However, the excess of particles leads to their undesired loss, and it also affects the flow properties of the Pickering emulsion.[19]

Another alternative is to introduce the particles in the droplets phase to minimize its waste. However, such an "Inside-Out" approach for the fabrication of Pickering emulsions is not suitable for most of "lab-on-chip" applications where the biochemical material inside droplets should replicate the one use in bulk experiments.[20]

In a microfluidic device, mixing is dominated by molecular diffusion due to the laminar flow conditions imposed by the small dimensions of the system. Hence, the large time scale observed for stabilization of droplets with particles in microchannels is most likely limited by diffusion-controlled transport of particles from the continuous phase to the droplet interface.[21,22]

Considering this, we propose a new design for droplet production where we aim to promote the adsorption of particles to the oil-water interface by increasing the confinement of particles near it: we take as reference the widely used "flow-focusing" droplet production method with cross-flow junction and we fashion a simple design modification where the continuous phase is separated –by function– in two streams, one for particle delivery and the second one for droplet production.[23,24]

We compare quantitatively both methods and demonstrate that our design modification reduces the droplet-stabilization time by one order of magnitude and at the same time significantly reduces the excess of particles required to stabilize the emulsion. Our experimental results therefore lead to design rules for a reliable emulsification using nanoparticles in microfluidics.

## 2. Results and Discussion

We first study the influence of experimental conditions on the stability of Pickering emulsions in microfluidics. We produce one microfluidic device design that allow us to test a standard flow-focusing production method (MI) and a modified version (MII) **(Figure 1a)**.

We produce monodispersed water-in-oil emulsions stabilized by particles with both configurations. For quantitative comparisons between MI and MII we consider as controlled parameters the ratio between the oil and aqueous flow rate $\lambda = Q_p/Q_w$, the fluorinated oil flow rate $Q_p$ and the concentration of fluorinated silica nanoparticles $c_p$ (w/v). For MII, $Q_p$ is divided in two streams: $Q_p^*$ which co-flows with $Q_w$ before the droplet production junction, and contains a concentration of fluorinated silica nanoparticles $c_p^*$ (w/v); and $Q_o$ which does not contain particles and joins the system at the cross-flow junction while diluting the particles concentration by a factor $c_p = c_p \frac{Q_p}{Q_p}$. Due to the laminar flow conditions imposed by the system dimensions, a long neck of $Q_p^*$ is formed in the incubation channel (as shown schematically in Figure 1a). Non-adsorbed particles will diffuse towards the particle-free fluorinated oil stream ($Q_o$) with enough given time. Additionally, we test two subcases of MII, by varying the fluorinated oil used for the stream $Q_o$: one subcase with HFE-7500 (MII$_{\eta\downarrow}$, 'low' viscosity) and another with the more viscous FC-40 (MII$_{\eta\uparrow}$, 'high' viscosity).

Downstream of the production junction, droplets are incubated for a time of $\tau = (Lh\omega_o)/(Q_p + Q_w)$ in a channel of fixed length ($L$ = 2 mm), width ($\omega_o$ = 100 μm) and depth ($h$ = 50 μm). During this time droplets remain separated from each other by the continuous phase until they reach a channel expansion (of width $6\omega_o$) where they are exposed to stochastic collisions and accelerations that favor coalescence **(Figure 1b)**.[25]

For each experimental condition, we measure the droplet volume ($V$) and monitor the impact of the experimental condition on coalescence. Quantitatively, we measure at least twenty thousand droplets and define the emulsion stability parameter against coalescence (α) which indicates the fraction of droplets that did not coalesce inside the expansion chamber, e.g. α > 0.999 indicates less than one coalescence event for one thousand droplets.[17]

In order to have a normalized reference of the particles available in the system, we estimate the concentration $c_{cp}$ (w/v) of particles needed to hexagonally close-pack a monolayer at the interface of a droplet of volume $V$ (if all the particles in the continuous phase were to adsorb) and we define particles in excess as $\varepsilon = c_p / c_{cp}$. The parameter $\varepsilon$ then indicates how many times in excess there are particles available in the continuous phase with respect to the needed to cover the droplets interface (**Figure 1c**).

## 2.1. Counterintuitive effect of droplet incubation time on Pickering emulsions stabilization

We investigate the effect of $\tau$ on droplet stability against coalescence for both production methods by fixing $\lambda$, $c_p$ and $c_p^*$ to a value of 2.5, 2 mg mL$^{-1}$ and 12 mg mL$^{-1}$ respectively. Since the incubation channel dimensions are fixed, we vary $\tau$ through the total flow rate ($Q_t = Q_p + Q_w$) from $Q_t = 70$ µL min$^{-1}$ ($\tau = 8.6$ ms) to $Q_t = 20$ µL min$^{-1}$ ($\tau = 30$ ms).

We observe an increase of $V$ with increasing $\tau$ for both MI and MII methods (**Figure 2b**). The droplet size is determined by the competition between the local shear stress, stretching interfaces and the resistance to deformation given by the capillary pressure, characterized by the dimensionless capillary number $Ca = \eta u/\gamma$.[26] For constant viscosity ($\eta$) and surface tension ($\gamma$), $Ca$ solely depends on the flow rate $Q_t$ ($=u\cdot\omega o\cdot h$). Consequently, $V$ decreases for larger $Ca$ (or $Q_t$). Droplet sizes depend on the two production methods. Using similar total oil flow rates, the droplets produced within M1 are consistently smaller droplets than within MII; in addition, higher oil viscosities produce smaller droplets. Qualitatively, this latter point is expected from the increase of the capillary number. Yet, the former observation indicates that the external oil flow only partly contributes to the shear at the interface, most likely because the oil flow at the junction did not have sufficient time to fully develop. This minor effect on the volume has however some important implication on other parameters: increasing V for fixed $\lambda$ and $c_p$ leads to a decrease in surface per unit volume: $c_{cp}$ decreases and $\varepsilon$ therefore increases. The increase of $\varepsilon$ with the incubation time observed in **Figure 2a**, is therefore consistent with the corresponding increase in droplet volume.

For these experiments, α is larger than 0.98 (**Figure 2c**) for all cases, corresponding to an emulsion with less than 2% of coalescence events. An excess of particles between 8 and 11 times the needed to cover the droplet interface prevents coalescence even for short incubation times (between 8 to 30 ms). In order to analyze the robustness of the stabilization process, we tested the system under more stringent conditions by measuring the stability for lower particle concentration. We reduce $c_p$ to 1 mg mL$^{-1}$ and measure α as a function of τ while monitoring $V$ and ε as previously.

We investigate the effect of τ on droplet stability against coalescence for both production methods by fixing λ, $c_p$ and $c_p^*$ to a value of 2.5, 1 mg mL$^{-1}$ and 12 mg mL$^{-1}$ respectively. As previously, we vary τ through the total flow rate ($Q_t = Q_p + Q_w$) from $Q_t$ = 70 μL min$^{-1}$ (τ = 8.6 ms) to $Q_t$ = 20 μL min$^{-1}$ (τ = 30 ms).

We observe again an increase of $V$ with increasing τ for both MI and MII methods **(Figure 2e)**. All the previous observations remain: the droplet size differs between the two production methods and with different viscosities and increasing τ correlates with an increase of ε as previously observed **(Figure 2d)**. However, under these more stringent conditions, α now displays values much smaller than 0.98 for MI **(Figure 2f)**, corresponding to unstable emulsions. In contrast, for both MII methods, α displays values larger than 0.99: an excess of particles between 4 and 5 is still able to prevent coalescence in our improved design. Remarkably, for case MII$_{\eta\uparrow}$, τ does not seem to have any influence on droplet stability against coalescence, obtaining highly stable emulsions (α > 0.999) even for τ values as low as 8.6 ms and presenting a small decrease on stability (α = 0.997) for τ > 30 ms.

A striking result we still need to highlight is the shape of α, which shows a decay as a function of τ and in some cases a non-monotonous behavior (Figure 2f). Naively, we would expect that increasing incubation time would yield more stable emulsions. However, our results show the opposite: droplet coalescence increases with increasing τ which indicates that under the studied conditions additional processes are dominating the dynamic stabilization of the Pickering emulsion against coalescence.

The decay of α observed for all cases coincides with a notable change in $V$, which seems to be related to a transformation of the droplet production mode from "dripping" to "squeezing" when $Q_t$ reaches 40 μL min$^{-1}$. This is observable on the production snapshots shown on **Figure 2g**, where the droplets produced transition from circular shape ($Q_t$ > 40 μL min$^{-1}$) to plug-like shape ($Q_t$ < 40 μL min$^{-1}$) inside the incubation channel.[27]

Furthermore, the behavior in MI at $c_p$ = 1 mg mL$^{-1}$ is informative: for low τ values, α increases with increasing τ, reaches a maximum at 15 ms ($Q_t$ = 40 μL min$^{-1}$) and then decreases with higher τ values. This behavior suggests a competition between two processes affecting droplet coalescence in MI at low $c_p$: for small droplets produced through dripping mode, longer τ allows further particle adsorption which yields more stable Pickering emulsions; while for larger droplets produced in squeezing mode, an additional process linked to the production mode and/or the size of the droplets produced dominates the Pickering emulsion stability.

In order to unravel the later mechanism, we want to summarize its possible origins: The stability of Pickering emulsions against coalescence relies on the adsorbed particles at the interface preventing the dispersed fluids from contacting each other.[28] When the droplet surface is not fully covered by absorbed particles, the drainage of the liquid films of the continuous phase between two approaching droplets becomes the rate-determining step for droplet contact and fusion.[29] An increase in droplet size increases the level of droplet confinement in the microfluidic device. Previous studies have found that geometrical droplet confinement results in additional hydrodynamic wall forces, which promotes coalescence due to an increased rate of film drainage between approaching droplets.[30,31] Additionally, smaller $Q_t$ increases the residence time of droplets in the coalescence chambers, giving more time for film drainage to occur.[32]

However, even though this effect could explain the decreasing stability with increasing droplet size for each individual case, it does not explain the differences between MI and MII: for $c_p$= 1 mg mL$^{-1}$, MII produces significantly more stable Pickering emulsions than MI, despite of producing always larger droplets than MI. This suggests that our improved method promotes particle adsorption.

From these experiments, we conclude that: for a fixed particle concentration, Pickering emulsions stability against coalescence in a microfluidic device is not necessarily improved by increasing the droplet incubation time. Additional processes –apparently related to droplet volume– are able to dominate the dynamic stabilization of Pickering emulsions. Additionally, the droplet production design MII bears a clear advantage for Pickering stabilization over the standard method of production MI, where the subcase MII$_{\eta\uparrow}$ yields the best result, producing Pickering emulsions with α > 0.999 for τ as low as 8.6 ms.

**2.2. The early stage of droplet production is key for Pickering emulsion stabilization**

In order to gain a better understanding on the role of droplet confinement on Pickering emulsion stability, we set new experimental conditions where we fix τ and modify V by varying λ. From our previous experiment, we select the data point where MI experienced maximum stability: τ is 15 ms ($Q_t$ =40 µL min$^{-1}$), $c_p$ is 1 mg mL$^{-1}$ (**Figure 3**, left panel) or 2 mg mL$^{-1}$ (Figure 3, right panel) and $c_p^*$ is 12 mg mL$^{-1}$.

We increase λ from 0.8 to 3. An increase of λ implies a decrease of the aqueous flow ($Q_w$) with respect to the particle dispersion flow ($Q_p$), which favors an increase of ε (more particles available per volume of water).

We then monitor α and $V$ as a function of ε. For both $c_p$ values and methods, $V$ decreases and α increases with increasing ε. The production of smaller droplets is a direct result of decreasing $Q_w$. Additionally, $V(MII_{\eta\downarrow}) > V(MII_{\eta\uparrow}) > V(MI)$ remains as a result of the inherit differences between the droplet production conditions highlighted in the previous section.

The increase of α is consistent with an increase of particles in excess, as well as a decrease of the droplet volume (less confined droplets). Interestingly, for ε values between 4 and 5, we obtain a range for which the two $c_p$ cases overlap presenting significantly different droplet size: for $c_p$ = 1 mg mL$^{-1}$ (Figure 3, left panel), $V$ varies between 150 and 280 pL; for $c_p$ = 2 mg mL$^{-1}$ (Figure 3, right panel), $V$ varies between 330 and 400 pL. We then compare the effect of droplet confinement on α for the same ε values.

For all cases, we obtained $V$ and α for ε = 4 by interpolating our data points using an exponential fit: for MI, the smaller droplet (167 pL) presents a smaller α value (0.899) in comparison to the significantly more confined case ($V$ = 332 pL, α = 0.982); for MII, in both cases α remains above 0.99, and we observe a minor decrease on stability for larger droplets at $c_p$ = 2 mg ml$^{-1}$. Because these two observations are at odds, we cannot explain them from the same physical origin, e.g. the effect of confinement: additional effects control the stabilization of the system. The higher stability for larger droplets observed for MI could be a result of the initial emulsification step: the dilution of particles in MI from $c_p^*$ to $c_p$ occurs off-chip. For $\lambda_1$ = 2.5 and $c_{p1}$ = 1 mg mL$^{-1}$ (Figure 3, central panel), the oil stream ($Q_{p1}$ = 28.6 µL min$^{-1}$) has a smaller local concentration of particles, than the similar data point at $\lambda_2$ = 1 and $c_{p2}$ = 2 mg mL$^{-1}$ ($Q_{p2}$ = 20 µL min$^{-1}$). We hypothesize that a higher local concentration of particles at the droplet formation junction is favoring particle adsorption, and consequently improving the emulsion stability.

This hypothesis is consistent with our data. Indeed, the higher local concentration of particles at the production explains the difference found so far between our new method MII and the standard method MI. For MII, the dilution from $c_p^*$ = 12 mg mL$^{-1}$ to $c_{p1}$ = 1 mg mL$^{-1}$ (or $c_{p2}$ = 2 mg mL$^{-1}$) occurs inside the chip at the flow-focusing point. $Q_{p1}^*$ (or $Q_{p2}^*$) co-flows with the aqueous phase when it reaches the flow-focusing point, where the stream of particle-free oil $Q_{o1}$ = 26.2 µL min$^{-1}$ (or $Q_{o2}$ = 16 µL min$^{-1}$) generates enough stress to break the aqueous stream into droplets. The local particle concentration during droplet formation is 12 mg mL$^{-1}$ (significantly more than method MI), increasing the adsorption of particles to water-oil interface. Furthermore, particle adsorption at the droplet-formation step would be aided by the shear-stress provided by the cross-flow –which also controls the droplet size– hinting at this

effect –rather than droplet confinement– as the one dominating the dynamic stabilization of Pickering droplets when the local particle concentration is constant.

Overall, we deduce that droplet formation is a critical step on Pickering emulsion stabilization (in a microfluidic chip): the adsorption of particles to the interface –aided by high local particle concentration and shear stress provided by the cross-flow– during droplet formation have a greater impact on stability than the bulk particle concentration and incubation time given to droplets downstream the microchannel.

**2.3. Improving the stabilization by hydrodynamically targeting particles to the interface**

We further test the influence of the local particle concentration and of the two production methods by designing a new set of experimental conditions: Here we aim to reduce the droplet size variations between production methods while keeping a substantial difference of the local particle concentration at the moment of droplet formation. We use the co-flow/flow-focusing combination for method MI **(Figure 4a)**, with $c_p$ in both oil streams: $Q_p^*$ (co-flow) and $Q_p$ (cross-flow). We fix the oil-aqueous ratio ($\lambda$ =2.7) as well as $Q_w/Q_p^*$ (4.28).

We vary $c_p^*$ (off-chip) between 1.3 and 6.5 mg mL$^{-1}$, obtaining $c_p$ values between 0.11 and 0.55 mg mL$^{-1}$. We monitor α as a function of ε for different incubation times (τ) by changing both L (from 5 to 20 mm) and $Q_t$ (from 40 to 80 µL min$^{-1}$). The droplet volume **(Figure 4b)** remains similar for both methods (MI and MII) and decreases with increasing $Q_t$, likewise our previous observations. For all cases, α increases with increasing ε **(Figure 4c)**. The increase of incubation length (L) does not affect the stability of the systems significantly. Only a minor increase on the stability is observed for the standard production method (MI) when L reaches 20 mm. When the total flow rate ($Q_t$) increases, an increase of α is observed for MI and MII$_{\eta\downarrow}$, when L = 5 mm. For the studied range of ε (from 1 to 6), MI could not reach α > 0.999 with incubation times as large as 150 ms; while MII$_{\eta\uparrow}$ reached it even for values as small as ε = 1.5 and τ = 18.75 ms. These results confirm the importance of the droplet formation step over final droplet stability. During this step most of the particles adsorb to the water-oil interface. This adsorption strongly depends on the local concentration of particles, as well as the viscous stress provided by the oil stream at the flow-focusing junction (which would be stronger for the more viscous oil (FC40) used in MII$_{\eta\uparrow}$). When insufficient number of particles are adsorbed during the formation step, very large incubation times are needed to obtain a stable system. In practice, our results bring a new design rule for droplet production in the presence of particles: maximizing the local particle concentration during droplet production, in addition to the viscous stress provided by the cross-flowed oil, is the most efficient mean to stabilize a Pickering emulsion in microfluidics.

## 3. Conclusion

In summary, we analyzed quantitatively the dynamics of stabilization of interfaces by nanoparticles in microfluidics. Our analysis revealed the crucial role of the flow profiles at the production junction on the stabilization kinetics. From our measurements, we determined design rules for the effective stabilization of Pickering emulsions in microfluidics. The critical parameters for an efficient stabilization are the nanoparticle concentration near the aqueous interface and the shear stress provided by the cross-flow during droplet formation. We control the former by adding a co-flow of particles dispersion with the aqueous phase, and the later by increasing the velocity or the viscosity of the cross-flow oil. These design rules allow us to significantly increase the droplet production throughput with a minimal amount of particle waste. Taking these guidelines, new designs could be implemented and adapted to the requirements of each technological application.

## 4. Experimental Section

### 4.1 Fluorinated silica nanoparticles synthesis

Pristine silica nanoparticles (Si-NPs) ($\delta$= 65 nm, **Figure 5a**) are synthesized in house following the protocol described by Hartlen et al. [33] Surface fluorination is performed as reported before: First, Si-NPs are dispersed in absolute ethanol (99%) at 10 mg ml$^{-1}$; Then, Ammonium hydroxide solution (30%) is added until a concentration of 2% v/v is reached; Finally, 1H,1H,2H,2H-perfluorooctyltriethoxysilane (PFOTES) is added to obtain a final concentration of 70 mg ml$^{-1}$.[13]

The reaction is left mixing at 250 rpm for two days. Then, the dispersion is centrifuged at 15000 rpm for 1 h during three washing cycles. After removing the supernatant, the particles are desiccated overnight in a vacuum chamber at room temperature. Finally, the particles are re-dispersed in HFE7500 (3M) using a sonication bath (15 min) followed by vortex mixing.

### 4.2 Determination of $c_{cp}$

The concentration needed ($c_{cp}$) to hexagonally close-pack a monolayer of particles at the droplet surface (S) was estimated: first, the number of particles needed at the interface ($n_{cp}$) was calculated assuming a 90° contact angle at the aqueous-oil interface and a hexagonal packing density of 0.9069.

$$n_{cp} = \frac{0.9069 S}{\pi \left(\frac{\delta}{2}\right)^2}$$

Then, $c_{cp}$ was calculated as

$$c_{cp} = \frac{\pi}{3} \frac{\delta^3}{2} \frac{\rho_p n_{cp}}{\lambda V}$$

Where $\delta$ is the particle diameter, and $\rho_p$ is the particle density, estimated as 0.165 g cm$^{-3}$ through titration with sodium hydroxide as introduced by Sears, (see supporting information).[34]

**4.3 Microfluidic device fabrication and experimental design**

The microfluidic device was designed and molded in PDMS using soft-litography techniques of replica molding of a SU-8 master with a pattern depth of $h$= 50 μm (**Figure 5b**).[5,35]

The PDMS was treated under oxygen plasma and bounded to a glass slide. The device channels were made hydrophobic by surface treatment with a commercial coating agent (Aquapel, PPG Industires). The microfluidic inlets were connected to syringes through Peek tubing of 0.75 mm inner diameter and the flow rates were controlled using syringe pumps (Nemesys, Cetoni).

The device was designed to test stability against droplet coalescence for two different production methods, which are achieved by selective puncture of the device inlets (**Figure 5c**): The first method (MI) consists in a standard flow-focusing production, where the particles are dispersed in the continuous phase (HFE-7500) and pre-diluted at the required concentration ($c_p$); in the second method (MII) the aqueous phase (deionized water) co-flows with a concentrated dispersion of particles ($c_p^*$) before reaching the droplet production junction where it meets (cross-flow) a stream of particle-free oil that dilutes $c_p^*$ down to $c_p$. Additionally, two different oils are tested for the cross-flow in MII: HFE-7500 (MII$_{\eta\downarrow}$), with a viscosity of 0.77 cSt and FC-40 (MII$\eta\uparrow$) with a viscosity of 2.2 cSt.

After production, the droplets are incubated in a channel of fixed width ($\omega$= 100 μm) which keeps them separated for a time proportional to a length (L). The incubation channel ends with an abrupt expansion of 600 μm that leads to a set of nine consecutive coalescence chambers of 800 μm length each (**Figure 5d**). This coalescence chamber was divided into sections with the aim of introducing larger random collisions between droplets, this helped to identify small differences on stability between emulsion conditions that seemed similarly stable at first (**Figure 5e**). However, even though effective as a reference to compare different systems stability against coalescence, reaching high stability ($\alpha$ > 0.999) does not guarantee full coverage of the droplets interface by particles. We empirically verified droplet coverage after emulsion collection by an optical readout (**Figure 5f**): upon drying, the volume of the droplet decreases. As the particles do not desorb from the surface, a buckling pattern is observed at the interface, characteristic of a layer of irreversibly adsorbed particles.[36,37]

For all the experiments, at least 1500 frames were recorded at a frame rate of 20 fps. The images were processed with Image-J software where the area of each droplet was extracted for the first coalescence chamber (Figure 5d (red)), and the apparent diameter (D) calculated as the median value of the droplet population distribution.[38] The droplet volume (V) and surface (S) was calculated using a nodoid shape approximation, which it has been shown to accurately describe a droplet confined in a microchannel.[39] The stability against coalescence (α) is the total number of non-coalesced droplets (*n(1)*), divided by the total amount of droplets described before by Baret *et al.*[17]

$$\alpha = \frac{n(1)}{\sum(i)}$$

Here *i* indicates the coalescence level, e.g. *i* = 3 corresponds to droplets which size is three times the original droplet size. We calculate α with data extracted from the last coalescence chamber (Figure 5d (green)).

**Supporting Information**

Supporting Information is available from the Wiley Online Library or from the author.

**Acknowledgements**

The authors thanks Dr. Pierre-Etienne Rouet for providing the pristine nanoparticles used in this study and Dr. Armand Roucher for making the TEM imaging of functionalized nanoparticles. This project has received funding from European Research Council (ERC) Seventh Framework Programme (2007-2013) ERC Grant Agreement 306385-SOFt Interfaces and from the European Research Council (ERC) under the European Union's Horizon 2020 research and innovation programme (grant agreement No 727480), from the 'Région Aquitaine' and from the French Government 'Investments for the Future' Programme, University of Bordeaux Initiative of Excellence (IDEX Bordeaux) (Reference Agence Nationale de la Recherche (ANR)-10-IDEX-03-02). J.-C. Baret acknowledges the support of the 'Institut Universitaire de France' (IUF). The technical support of the chemistry facility at the Centre de Recherche Paul Pascal (Pessac, France) is warmly acknowledged.

References

[1] R. Seemann, M. Brinkmann, T. Pfohl, S. Herminghaus, *Rep. Prog. Phys.,* **2012**, 75, 16601-2012.
[2] A. Theberge, F. Courtois, Y. Schaerli, M. Fischlechner, C. Abell, F. Hollfelder, W. Huck, *Angew. Chem. Int. Edit.,* **2010**, 49, 5846-5868.
[3] A. M. Klein, L. Mazutis, I. Akartuna, N. Tallapragada, A. Veres, V. Li, L. Peshkin, D. A. Weitz, M. W. Kirschner, *Cell,* **2015**, 161, 1187-1201.
[4] T. Beneyton, I. P. M. Wijaya, P. Postros, M. Najah, P. Leblond, A. Couvent, E. Mayot, A. D. Griffiths, A. Drevelle, *Sci, Rep-UK.,* **2016**, 6, 6.
[5] J.-C. Baret, *Lab Chip,* **2012**, 12, 422-433.
[6] J.-W. Kim, D. Lee, H. C. Shum, D. A. Weitz, *Adv. Mater.,* **2008**, 20, 3239-3243.
[7] P. Gruner, B. Riechers, L. A. C. Orellana, Q. Brosseau, F. Maes, T. Beneyton, D. Pekin, J.-C. Baret, *Curr. Opin. Colloid In.,* **2015**, 20, 183-191.
[8] M. Pan, L. Rosenfeld, M. Kim, M. Xu, E. Lin, R. Derda, S. K. Y. Tang, *ACS Appl. Mater. Inter.,* **2014**, 6, 21446-21453.
[9] M. Pan, M. Kim, L. Blauch, S. K. Y. Tang, *RSC Adv.,* **2016**, 6, 39926-39932.
[10] Y. Gai, M. Kim, M. Pan, S. K. Y. Tang, *Biomicrofluidics,* **2017**, 11.
[11] I. Platzman, J.-W. Janiesch, J. P. Spatz, *J. Am. Chem. Soc.,* **2013**, 135, 3339-3342.
[12] S. Fouilloux, F. Malloggi, J. Daillant, A. Thill, *Soft Matter,* **2016**, 12, 900-904.
[13] L. A. Chacon, J. C. Baret, *J. Phys. D Appl. Phys.,* **2017**, 50, 39LT04.
[14] M. Pan, F. Lyu, S. K. Y. Tang, *Anal. Chem.,* **2015**, 87, 7938-7943.
[15] C. Priest, M. D. Reid, C. P. Whitby, *J. Colloid Interf. Sci.,* **2011**, 363, 301-306.
[16] X. Yao, Z. Liu, M. Ma, Y. Chao, Y. Gao, T. Kong, *Small,* **2018**, 14, 1802902.
[17] J.-C. Baret, F. Kleinschmidt, A. E. Harrak, A. D. Griffiths, *Langmuir,* 2009, 25, 6088-6093.
[18] B. Riechers, F. Maes, E. Akoury, B. Semin, P. Gruner, J.-C. Baret, *P. Natl. Acad. Sci. USA,* **2016**, 113, 11465-11470.
[19] A. B. Subramaniam, M. Abkarian, H. A. Stone, *Nature Materials,* **2005**, 4, 553-556.
[20] Z. Nie, J. I. Park, W. Li, S. A. F. Bon, E. Kumacheva, *J. Am. Chem. Soc.,* **2008**, 130, 16508-16509.
[21] O. S. Deshmukh, D. Ende, M. C. Stuart, F. Mugele, M. H. G. Duits, *Adv. Colloid Interfac.,* **2015**, 222, 215-227.
[22] H. A. Stone, A. D. Stroock, A. Ajdari, *Annu. Rev. Fluid Mech.,* **2004**, 36, 381-411.
[23] S. L. Anna, N. Bontoux, H. A. Stone, *Appl. Phys. Lett.,* **2003**, 82, 364-366.
[24] A. M. Pit, M. H. G. Duits, F. Mugele, *Micromachines,* **2015**, 6, 1768-1793.
[25] N. Bremond, A.R. Thiam and J. Bibette, *Phys. Rev. Letters,* **2008,** 100, 024501
[26] L. Shang, Y. Cheng, Y. Zhao, *Chemical Reviews,* **2017**, 117, 7964-8040.
[27] M. De Menech, P. Garstecki, F. Jousse, H. A. Stone, *J. Fluid Mech.,* **2008**, 595, 141–161.
[28] D. E. Tambe, M. M. Sharma, *J. Colloid Interf. Sci.,* **1993**, 157, 244-253.
[29] T. Krebs, K. Schroën, R. Boom, *Soft Matter,* **2012**, 8, 10650.
[30] D. Chen, R. Cardinaels, P. Moldenaers, *Langmuir,* **2009**, 25, 12885-12893.
[31] P. D. Bruyn, R. Cardinaels, P. Moldenaers, *J. Colloid Interf. Sci.,* **2013**, 409, 183-192.
[32] Q. Zhou, Y. Sun, S. Yi, K. Wang, G. Luo, *Soft Matter,* **2016**, 12, 1674-1682.
[33] K. D. Hartlen, A. P. T. Athanasopoulos, V. Kitaev, *Langmuir,* **2008**, 24, 1714-1720.
[34] G. W. Sears, *Anal. Chem.,* **1956**, 28, 1981-1983.

[35] Y. Xia, G. M. Whitesides, *Angew. Chem. Int. Edit.,* **1998**, 37, 550-575.
[36] S. S. Datta, H. C. Shum, D. A. Weitz, *Langmuir,* **2010**, 26, 18612-18616.
[37] N. Tsapis, E. R. Dufresne, S. S. Sinha, C. S. Riera, J. W. Hutchinson, L. Mahadevan, D. A. Weitz, *Phys. Rev. Lett.,* **2005**, 94, 018302.
[38] J. Schindelin, I. Arganda-Carreras, E. Frise, V. Kaynig, M. Longair, T. Pietzsch, S. Preibisch, C. Rueden, S. Saalfeld, B. Schmid, J.-Y. Tinevez, D. J. White, V. Hartenstein, K. Eliceiri, P. Tomancak, A. Cardona, *Nat. Methods,* **2012**, 9, 676-682.
[39] J. Lim, O. Caen, J. Vrignon, M. Konrad, V. Taly, J.-C. Baret, *Biomicrofluidics,* **2015**, 9, 034101.

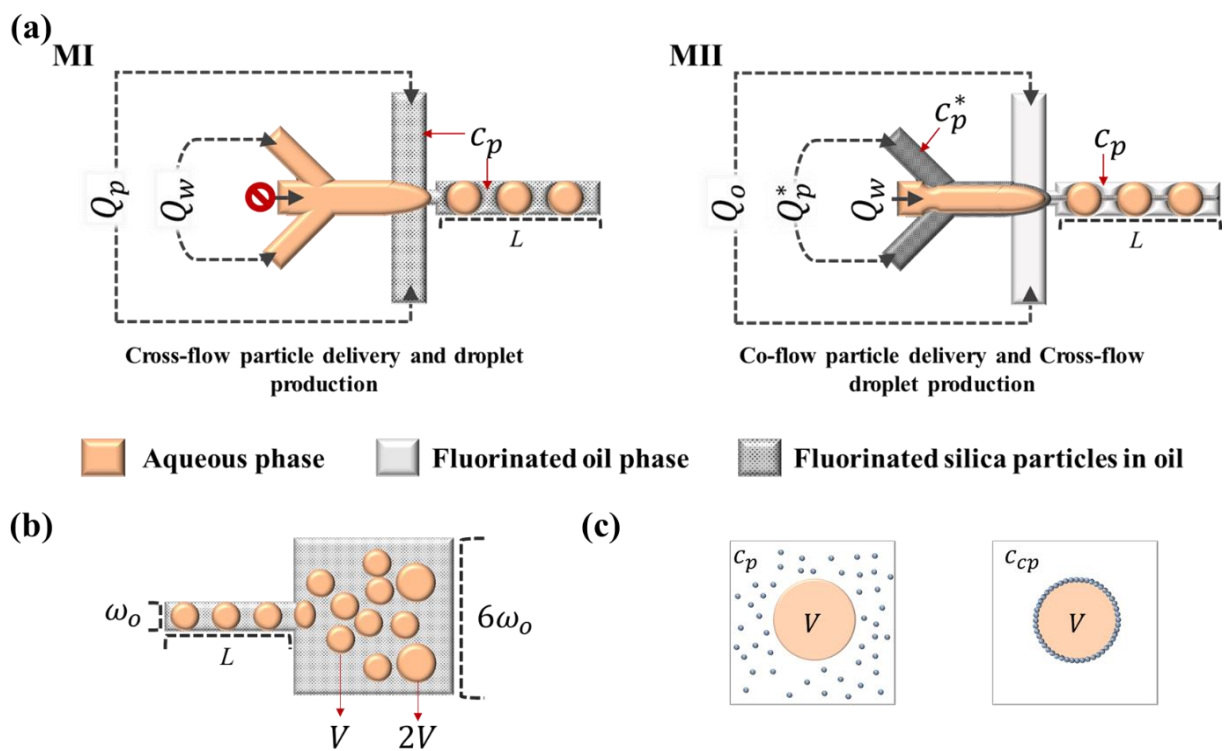

**Figure 1.** (a) Schematic representation of a standard droplet production method (MI) and a new droplet production design (MII). (b) Incubation channel and coalescence chamber dimensions. (c) Schematic representation of particle concentration ($c_p$) vs. concentration needed ($c_{cp}$) to hexagonally close-pack a monolayer of particles at the droplet interface.

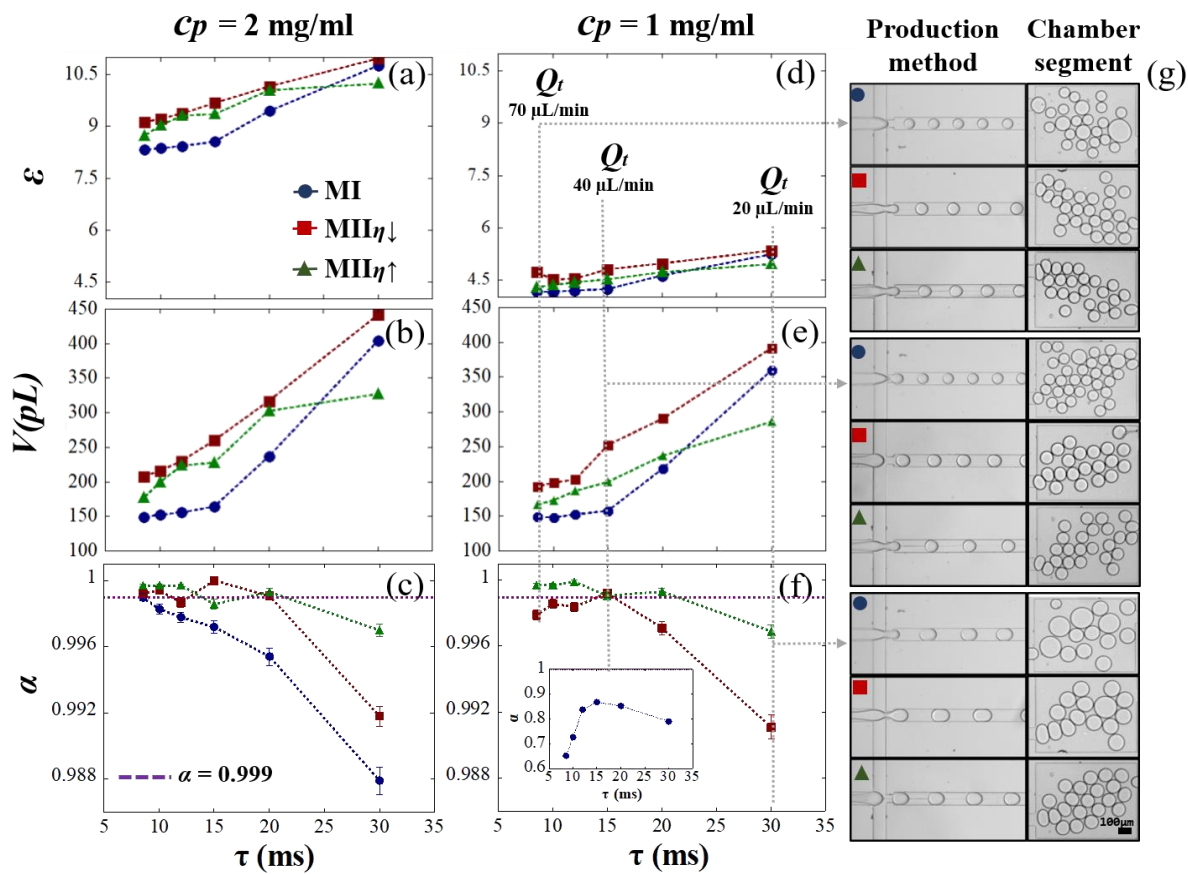

**Figure 2.** Comparison of methods MI and MII ($Q_o$ being ($\eta\downarrow$) HFE7500 and ($\eta\uparrow$) FC40) on the effect of $\tau$ on $\alpha$, V (droplet volume) and $\varepsilon$. $\lambda$, $L$ and $c_p^*$ are kept constant, $c_p$ is also kept constant at: (a-c) 2 mg/ml and (d-g) 1 mg/ml, where we show snapshots of droplet production and the last segment of coalescence chambers for three data points ($Q_t$= 70, 40 and 20 µL min$^{-1}$). The error bars in c and f are calculated as the square root of the number of coalesced event detected. Due to the significant differences in the values of a, the data for MI are shown as an inset in f.

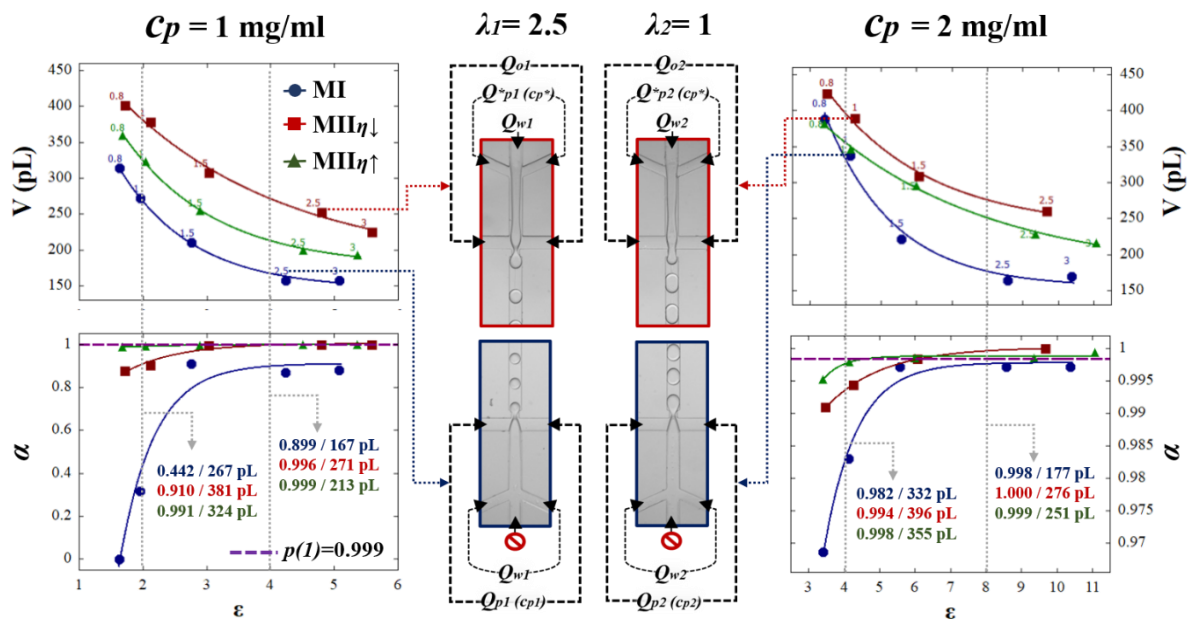

**Figure 3.** Comparison of methods MI and MII ($Q_o$ being ($\eta\downarrow$) HFE7500 and ($\eta\uparrow$) FC40), for a fixed incubation time ($\tau$ = 15 ms). Stability ($\alpha$) is monitored as a function of $\varepsilon$, which is varied along with $V$ by increasing $\lambda$ from 0.8 to 3. $\alpha$ and $V$(pL) values are calculated for $\varepsilon$ = 2, 4 and 8 (an exponential decay fit is used to interpolate the data). $c_p$ is kept constant at: 1 mg ml$^{-1}$ (left, production snapshot for MI and MII$_{\eta\downarrow}$ at $\lambda$ = 2.5) and 2 mg ml$^{-1}$ (right, production snapshot for MI and MII$_{\eta\downarrow}$ at $\lambda$=1).

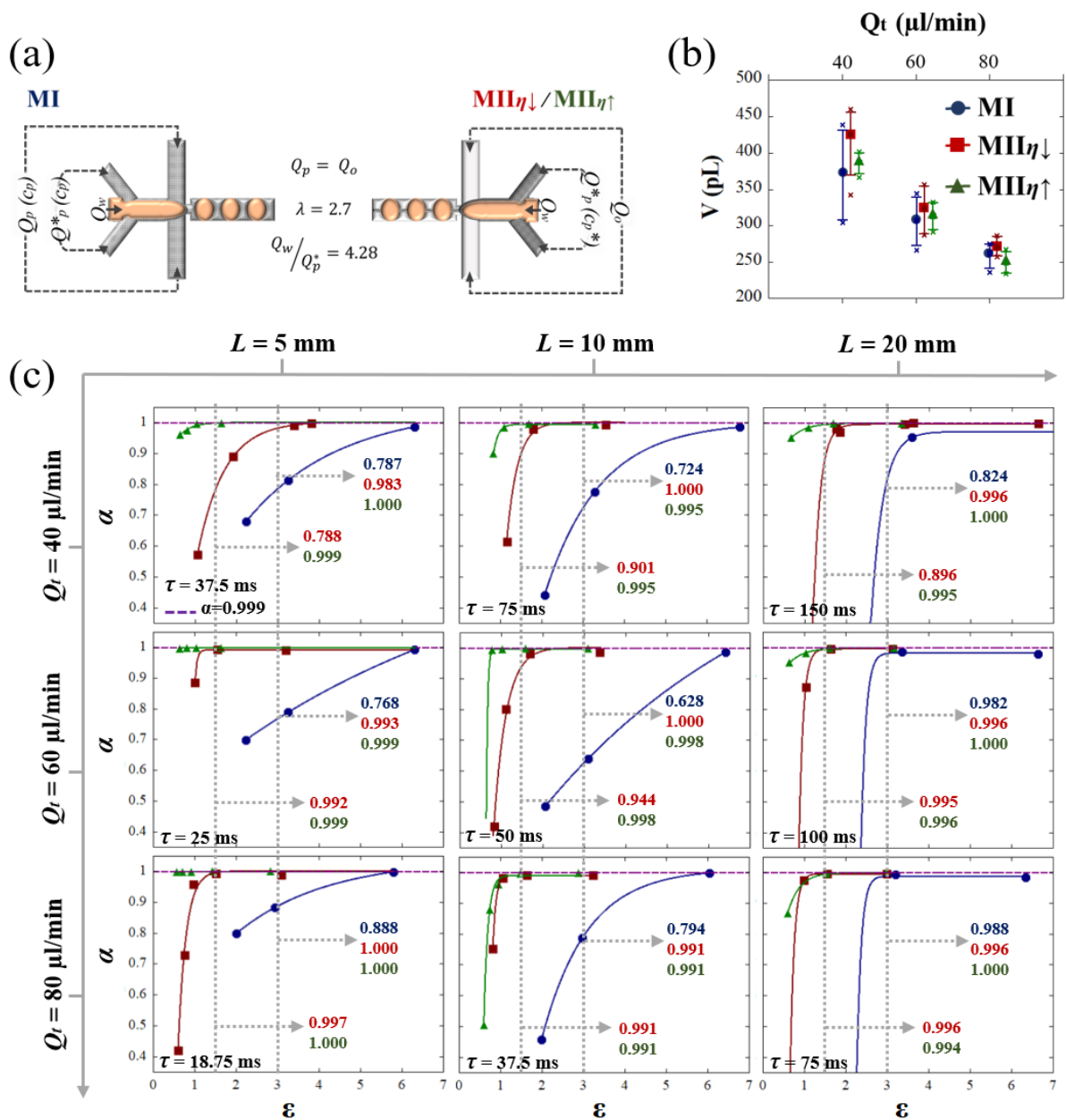

**Figure 4.** (a) Schematic representation of experimental conditions for method MI and MII ($Q_o$ is ($\eta\downarrow$) HFE7500, ($\eta\uparrow$) FC40). (b) $V$ variation as a function of $Q_t$ for MI and MII, the error bars represent the standard deviation and the crosses the maximum and minimum values. (c) α variation as a function of ε for $L$ between 5 and 20 mm and $Q_t$ between 40 and 80 μL min$^{-1}$.

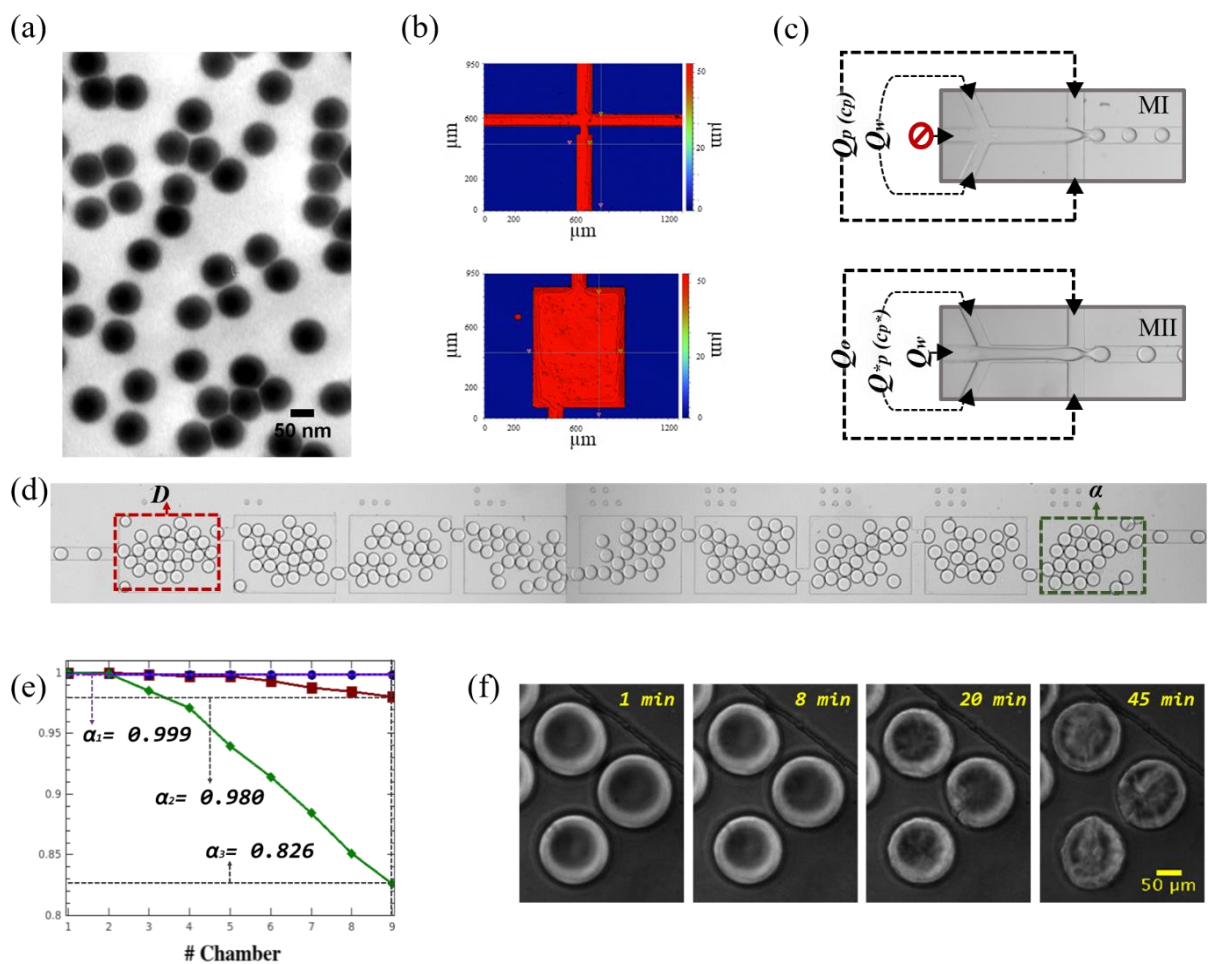

**Figure 5.** (a) TEM imaging of Fluorinated silica nanoparticles. (b) Profilometer measurements for the determination of the SU-8 master pattern depth. (c) Operation modes for droplet production with the same microfluidic pattern: for method MI, the central inlet is not pierced. (d) Coalescence chamber design divided into nine segments, droplet size is extracted from the first chamber, α value is extracted from the last chamber. (e) Stability parameter calculated in each chamber section. (f) Empirical observation of buckling pattern at the collected droplets interface upon drying.